\documentclass[sigconf]{acmart}

\setcopyright{none}
\renewcommand\footnotetextcopyrightpermission[1]{}

\usepackage{fancyhdr}
\usepackage{balance}

\usepackage{xcolor}
\usepackage{xspace}
\newcommand\system{\textsc{P4rrot}\xspace}

\begin{document}

\title{P4RROT: Generating P4 Code for the Application Layer}

\author{Csaba Györgyi}
\affiliation{
	\institution{E\"{o}tv\"{o}s Lor\'{a}nd University, Hungary \\ University of Vienna, Austria}
}
\email{gycsaba96@inf.elte.hu}

\author{S\'{a}ndor Laki}
\affiliation{
	\institution{ELTE E\"{o}tv\"{o}s Lor\'{a}nd University, Hungary}
}
\email{lakis@inf.elte.hu}

\author{Stefan Schmid}
\affiliation{
	\institution{TU Berlin, Germany \\ University of Vienna, Austria}
}
\email{stefan.schmid@tu-berlin.de}

\begin{abstract}
Throughput and latency critical applications could often benefit of performing computations close to the client. To enable this, distributed computing paradigms such as edge computing have recently emerged.
However, with the advent of programmable data planes, computations cannot only be performed by servers but they can be offloaded to network switches. Languages like P4 enable to flexibly reprogram the entire packet processing pipeline.
Though these devices promise high throughput and ultra-low response times, 
implementing application-layer tasks in the data plane programming language P4
is still challenging for an application developer who is not familiar with networking domain. 
In this paper, we first identify and examine obstacles and pain points one can experience when offloading server-based computations to the network. 
Then we present \system,
a code generator (in form of a library) which allows to overcome these limitations by providing a user-friendly API to describe computations to be offloaded.
After discussing the design choices behind \system, we introduce our proof-of-concept implementation for two P4 targets: Netronome SmartNIC and BMv2.
\end{abstract}






\keywords{offloading, programmable dataplanes, code generation, P4}

\maketitle

\section{Introduction}

The compute infrastructure is becoming increasingly
distributed, offering multiple locations to serve requests and execute
applications. 
This introduces interesting opportunities for spatial optimization:
by bringing computation and data storage closer to the requester (or client),
response times and bandwidth can often be greatly improved. 

Early examples of this paradigm are the distributed domain name system and the 
content distributed networks created in the late 1990s~\cite{dilley2002globally}. More recent examples
include the edge computing paradigm as well as the trend towards in-network 
computing.

This paper is motivated by a novel offloading opportunity introduced by
software-defined networks~\cite{6994333}, and in particular programmable dataplanes~\cite{csur21}. 
Programmable networks have recently received much attention both in academia
and industry, for their support of fast networking innovations: 
while developing new semiconductors is a time-consuming and expensive process,
a programmable data plane provides an efficient and flexible way to support
new protocols and new requirements. The dataplane programming language P4 
(Programming Protocol-Independent Packet Processors) is independent of the forwarding hardware design, and relies on a compiler specific to the hardware
(e.g., SmartNICs, NetFPGAs, programmable ASICs), 
thus allowing the fast implementation of new protocols and networking algorithms.
Over the last years, several interesting use cases for programmable dataplanes
have been demonstrated that are related to resilience, security, resource allocation, among others~\cite{csur21}. Like how people started using GPUs for non-graphics-related use cases (e.g., crypto mining and deep learning), many research projects focus on how programmable dataplanes could be used for executing application layer tasks (e.g., emergency stops, robot control, or even key-value stores~\cite{jin2017netcache,brazilokcikke,p4pid}).  

P4 has proven useful for implementing application-layer tasks, but even simple functionalities can quickly result in complex software projects. This is due to the limitations of the underlying devices and the fact that P4 was not designed to support this kind of computations. For example, consider the following scenario which highlights some of the challenges that developers can face when implementing L7 logic.
Suppose that in order to precalculate an aggregated value to reduce the CPU load, a programmer wants to calculate and insert an additional integer at the beginning of the UDP payload. This seemingly simple task requires to parse the usual headers, adjust the total length field in the IPv4 and UDP headers, recalculate the checksum, and implement a simple static forwarding. Even if the programmer already wrote similar and relevant code in the past (e.g., a ring buffer made of registers), it is hard to reuse this implementation since P4 does not easily support to encapsulate such a high-level abstraction. Along the way, the programmer might further need to use workarounds and different tricks to avoid resource limitations and compiler bugs. Whether calculations are peformed with a table or if-else statements may appear to be an implementation detail, but is actually a design decision from a P4 point of view, requiring additional effort. If the programmer then wants to further extend the functionality, e.g., by inserting a second integer, she must modify the existing code in multiple different places. Also testing is complex, and may require a complete pipeline with whole packets, as there is typically no good way to unit test implementation parts. 

We argue that automatic code generation can greatly simplify
implementing application-layer tasks if we narrow down the scope of target features.
By application-layer tasks, we mean functionalities concerned about the payload rather than other networking layers.
Against this backdrop, our main \textbf{\emph{contribution}} in this paper is 
\system, a code generator (in form of a library) for P4, 
to support and speed up the offloading process. 
With \system, the developer can describe the application layer logic
using our (Python3) library~\cite{p4rrot-github}  and generate the equivalent P4 code.
\system~does not require a new programming language, which also simplifies the 
adoption of new P4 features.
In this paper, we discuss the different design aspects and report on an 
example implementation of \system~for 
the BMv2 (Behavioral Model v2)~\cite{bmv2} and Netronome NFP (Netronme Flow Processor)~\cite{nfp} targets, written 
in Python3.
We chose the BMv2 and Netronome NFP targets because they both use the V1Model architecture. The BMv2 makes our implementation easy to run, while the NFP smartNIC is a sensible target, it seems better suited for server offloading than rigid ASICs.

\begin{figure}
    \centering
    \includegraphics[width=0.45\textwidth]{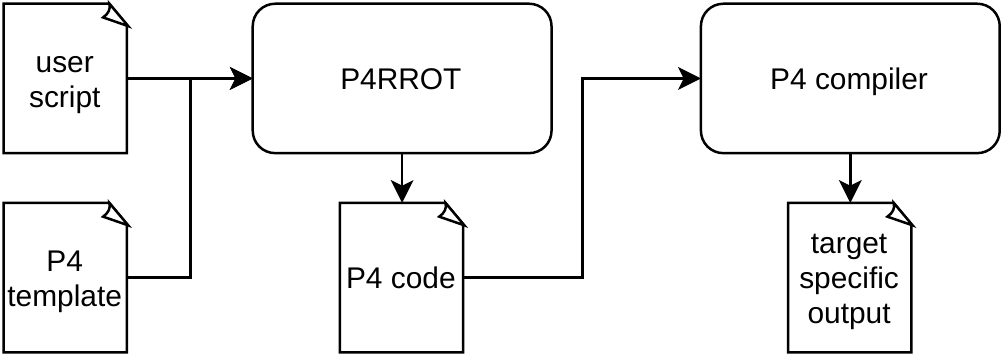}
    \caption{The high-level overview of the use of \system.}
    \label{fig:system}
\end{figure}

Fig. \ref{fig:system} illustrates the architecture of our proposed solution.

\section{Goals and Requirements for P4RROT}
\label{sec:requirements}

This section lists the main principles the design and implementation of \system~follows, divided into two categories. First, we highlight the features that overcome the previously described pain points. Second, we describe additional requirements improving the adaptability of our solution.

\subsection{Painkillers}
\label{subsec:painkillers}

An important decision underlying \system's design is that we limit our target programs to the set of offloaded application layer logic, thus making the design simpler in many ways. Having to deal only with a limited amount of commands and data types (e.g., a 3-bit long field does not make much sense in this context), we can make powerful assumptions and automate many tasks. 
Simplicity and structure allow the developers to keep their work compact and prevent code fragmentation.
Furthermore, we require that implementation details should remain implementation details, preferably in the form of ``hints''. For example, checking the equality with an if-else statement or using a table can be simply an optional parameter.
Our solution should further be able to encapsulate complex, often used logic, and hide data structures and algorithms (like a ring buffer made of registers) behind a straightforward API. Adding similar extensions should also be easy.
Last but not least, once we fully or partially described the business logic, we want to simulate the behaviour of the application layer without even generating the P4 code. Being able to account for overflows and other low-level details is crucial.

\subsection{Adaptability}
\label{subsec:adaptibility}

The code generator should leverage an existing and well-known language in an easy-to-understand way, so programmers can work with a familiar syntax and are productive quickly. 
Generally, the generated P4 code should be human-readable and reasonably easy to understand. The programmer should still have the opportunity to override the code generator's decision, e.g., for further performance optimizations. 
The generator only performs simple semantic checks, and for instance, does not override constant values. By avoiding complex verifications, the implementation of possible extensions is simplified. We also note that target-specific checks (e.g. resource constraints) might be even impossible to implement because of legal restrictions.
Also error messages should be easy to understand and point to the line where the programmer made the mistake. Since \system~is essentially a library that allows programmers to describe the abstract syntax tree (AST) of the offloaded solution by defining a complex object, the semantic checks are run after adding each and every node. By doing so, the generator library can raise an exception at any line as early as possible.

\section{Architecture and design}
\label{sec:design}

This section describes the internal working of \system's code generator through the 
main components of an offloading project. For a better understanding, we provide a simple example first. Then we explain the main design decisions and architectural elements behind the scenes.

\subsection{A simple example}

\begin{figure}
    \centering
     \includegraphics[width=0.45\textwidth]{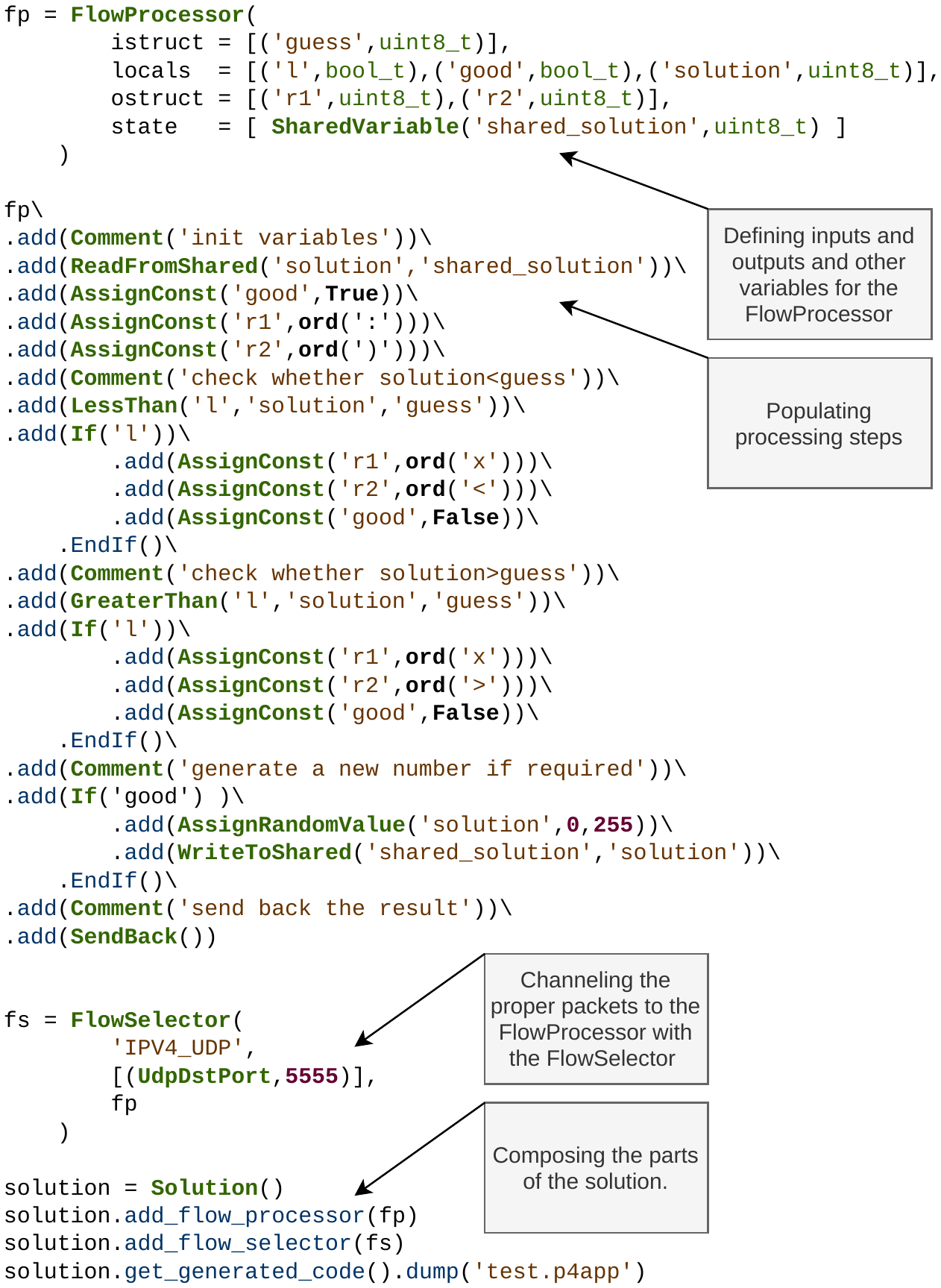}
    \caption{Sample usage of P4RROT implementing a number guessing game}
    \label{fig:numberguessing}
\end{figure}

To illustrate the various features and the style of an offloading project,
let us consider a simple example. 
Fig. \ref{fig:numberguessing} shows the implementation of a simple number guessing game. Although it is not particularly useful, it provides an easy-to-understand task that does not require any background knowledge. The client has to figure out a randomly generated number. After each guess, there are three possible outcomes: the solution is lower than the provided number, the solution is greater than the provided number, or the client wins (and a new number is generated). 

First, we create a \verb|FlowProcessor| in a declarative input-output style. The input is a single byte representing the client's guess, and the output is two bytes (treated as characters in the later stages). Additionally, we use some local variables and a shared variable to store the correct solution.
After that, we populate the processing steps with various commands to define our application-level algorithm. We assume that the client has the correct answer, and then we change it if the guess is less or greater than the right solution in the \verb|SharedVariable|. In the end, we send back the packet where it came from with a single \verb|Command|.
Using a \verb|FlowSelector|, we also need to define which packets should be processed by the previously described \verb|FlowProcessor|. 
In the end, we assemble the parts on a single object and generate the P4 code using the provided template.

\begin{figure}
    \centering
    \includegraphics[width=0.45\textwidth]{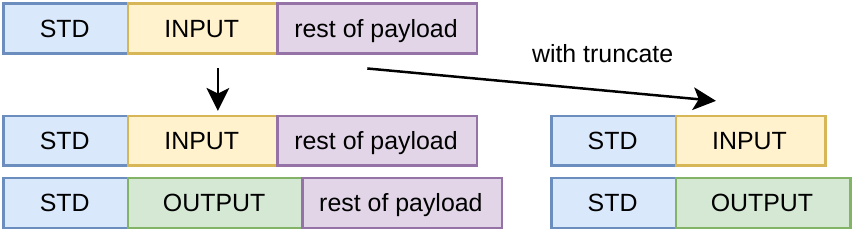}
    \caption{The different ways the user can transform a packet.}
    \label{fig:pkt-transf}
\end{figure}

\subsection{Processors}

\verb|FlowProcessor| objects are the essence of the offloaded solution describing the application-layer calculations. 
During the instantiation, the programmer defines the input and the optional output structures. By default, the code generator sets the input header invalid and the output header valid, thus leaving room for modifying the packet structure. Moreover, if the \verb|truncate| extern or similar functionality is available, the remaining payload can also be removed. If the output structure is not defined, the original input header is not invalidated. Fig. \ref{fig:pkt-transf} depicts the different ways the user can transform a packet. The use of local variables, temporary headers and, stateful elements are also possible. 

A \verb|FlowProcessor| contains a \verb|Block| object which encapsulates a sequence of \verb|Command|s. \verb|Command|s are responsible for describing different operations performed on the packets. They can vary from low-level casting to high-level data structure manipulations. Besides the necessary input and output variables, they might provide different implementation hints using optional variables. 

\begin{figure}
    \centering
    \includegraphics[width=0.45\textwidth]{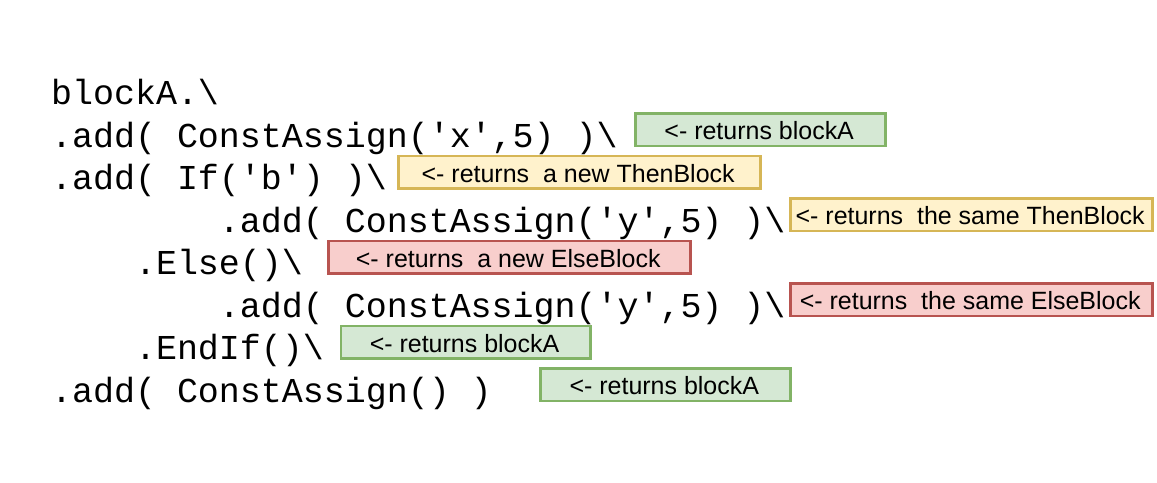}
    \caption{An example explaining the internal workings of the  if-else statement's implementation.}
    \label{fig:ifelse}
\end{figure}

When a new  \verb|Command| is added to a \verb|Block|, it returns itself or another \verb|Block|. Thanks to this design, it is possible to describe algorithms reasonably intuitively, similar to the JavaScript Promises or the LINQ library in C\#.  Comfortably defining if-else statements is also possible. Once we add an \verb|If| command, we return a \verb|ThenBlock| (which inherits from the \verb|Block| class). The \verb|ThenBlock| instance can return an \verb|ElseBlock| using its \verb|Else| method. Finally, the \verb|ElseBlock|'s \verb|EndIf| returns the original parent \verb|Block|. Fig.~\ref{fig:ifelse} depicts 
a simple example. We define \verb|Switch| statements and \verb|Atomic| blocks similarly. 

Leveraging this design, the code generator library can provide simple semantic checks upon adding \verb|Command|s and simulate the behavior of a \verb|FlowProcessor|.

\subsection{Selectors}

\begin{figure}
    \centering
    \includegraphics[width=0.45\textwidth]{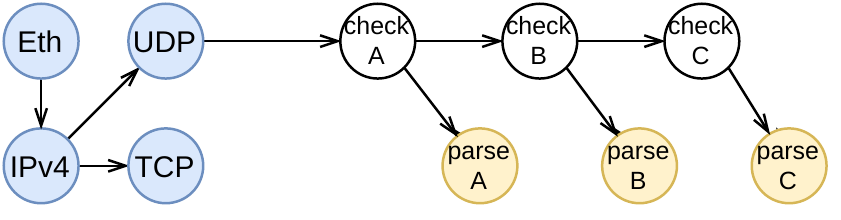}
    \caption{A simple parser-chain.}
    \label{fig:paser-chain}
\end{figure}

To define what kind of packets are processed in the offloaded solution, 
the user script can create \verb|FlowSelectors|: simple objects using the code generator library. 
First, the \verb|FlowSelector| defines the underlying standard protocols with a single constant (e.g. \texttt{IPV4\_UDP}). Since one might have different selectors for the same underlying protocols, the code generator organizes the P4 parser states in a chain-like structure (see Fig.~\ref{fig:paser-chain}). If the condition of the selector is met, then we extract the application-layer data, otherwise we check the next selector in a different state. If there is no next selector, we proceed to the next desired state, e.g.~\verb|accept|.
To deal with unused, empty chains, one can use \texttt{\#define} pragmas. The P4 template always expects an empty chain unless a particular macro is defined. The generator can easily insert this \texttt{\#define} pragma into the generated parser code.

The programmer can define the selection criteria using a list of pairs. Each pair consists of a field name and a required value. The used field can be a standard field of a standard header (e.g., UDP destination port) or the member of a user-defined struct describing the beginning of the application data. The latter one leverages the lookahead capabilities of the P4 parser.

\subsection{Templates} 

Templates serve as a static starting P4 code for an offloading project. It can be used to describe the parsing of standard headers or implement basic forwarding rules.
The template only references the generated code parts using the \texttt{\#include} preprocessor pragma. Theoretically, the developer can use as many extra generated files as she wishes. However, at least one include pragma in the custom headers, parsing, header-struct body, ingress declarations, and ingress \texttt{apply} block seem necessary.
A template must also provide an interface for the code generator by defining certain metadata variables and macros. For example, we maintain the number of added and removed bytes using metadata variables and provide workarounds for the atomic block using macros. One might use macros for every purpose, thus making the template codes more flexible and more complex at the same time.

\section{Additional Related work}
\label{subsec:relatedwork}

Generating lower-level code for a specific use case from a high-level language 
is a common theme for simplifying application development and used in many contexts.
For example, TensorFlow~\cite{abadi2016tensorflow} allows Python users to create computation graphs and run them on GPU, and Keras~\cite{ketkar2017introduction} allows to define and use neural networks using a simple API. A similar networking related example is MoonGen\cite{emmerich2015moongen} which facilitates performance measurements using DPDK and Lua. However, these do not focus on P4.

We are also not the first to present a code generator which outputs 
P4 code. For example, Graph-To-P4~\cite{zaballa2019graph} is a boilerplate code generator for parser graphs. Also different packet filtering solutions convert logical expressions (concerning application data) to P4 code, such as CAMUS~\cite{jepsen2020forwarding} and FastReact~\cite{vestin2018fastreact}. LUCID~\cite{sonchack2021lucid} is an entirely new language meant to implement control plane functionality in P4 data planes.
However, these examples do not focus on offloading arbitrarily defined application tasks. To best of our knowledge, we are the first to provide an interface that is capable of describing general application-layer features.

\section{Discussion and Future Work}

We presented a code generator, \system, which allows to provide a familiar and straightforward interface to the P4 programmer, hence simplifying application
offloading.
\system's simple interface is possible by narrowing down the scope to offloading application functionalities.
 We note that our approach is not limited to the Python language and 
the same design can be used for other high-level languages, e.g. Java, C\#. 

While \system~can already be useful in its current form, our project is still in an early stage. Due to the limited number of implemented commands and stateful elements, \system's  expressiveness is narrow at the moment. \system~is open source and we can imagine even complex behaviors defined in tens of lines of code, using bloom-filters, heaps, floating-point operations and other high-level abstractions. We also plan to add extra functionalities to the existing system, including a built-in way for the interaction between the data plane and the non offloaded server components. 

Furthermore, although we so far considered only the Netronome NFP and BMv2 as a target, our approach is not architecture-specific. From our experiences, TNA developers can greatly benefit from quickly switching implementation alternatives with optional parameters (hints).

\system~may also open a business opportunity for companies. We can imagine scenarios in which the templates and some additional extensions are proprietary and in which the end-users write python scripts. After the code generation, the P4 code is automatically compiled and loaded to the company's device.

\begin{acks}
Research supported by the German Bundesministerium für Bildung und Forschung (BMBF) project, 
6G-RIC: 6G Research and Innovation Cluster, 2021-2025.
\end{acks}

{ \balance
{
	\bibliographystyle{ACM-Reference-Format}
    \bibliography{./bibliography/IEEEabrv,./bibliography/IEEEexample}
%
%
}
}

\end{document}